\documentclass{llncs}

\usepackage{color}
\usepackage{booktabs}
\usepackage{times}
\usepackage{subfig}
\usepackage{verbatim}
\usepackage{makeidx}
\usepackage{amsmath}
\usepackage[nocompress]{cite}
\usepackage[utf8]{inputenc}

\usepackage{ifpdf}
\ifpdf
\usepackage[pdftex]{graphicx}
\usepackage{epstopdf}
\else
\usepackage{graphicx}
\fi

\begin{document}

\pagestyle{headings}

\title{An Empirical Study of the Repair Performance of Novel Coding Schemes for Networked Distributed Storage Systems}

\author{Lluis Pamies-Juarez\inst{1} \and Fr\'ed\'erique Oggier\inst{1} \and Anwitaman Datta\inst{2}}
\institute{School of Physical and Mathematical Sciences \and School of Computer Engineering\\Nanyang Technological University, Singapore}

\maketitle


%
%
\begin{abstract}
Erasure coding techniques are getting integrated in networked distributed storage systems as a way to provide
fault-tolerance at the cost of less storage overhead than traditional replication. Redundancy is maintained over time
through repair mechanisms, which may entail large network resource overheads. In recent years, several novel codes
tailor-made for distributed storage have been proposed to optimize storage overhead and repair , such as Regenerating
Codes that minimize the per repair traffic, or Self-Repairing Codes which minimize the number of nodes contacted per
repair. Existing studies of these coding techniques are however predominantly theoretical, under the simplifying
assumption that only one object is stored. They ignore many practical issues that real systems must address, such as
data placement, de/correlation of multiple stored objects, or the competition for limited network resources when
multiple objects are repaired simultaneously.  This paper empirically studies the repair performance of these novel
storage centric codes with respect to classical erasure codes by simulating realistic scenarios and exploring the
interplay of code parameters, failure characteristics and data placement with respect to the trade-offs of bandwidth
usage and speed of repairs.
\end{abstract}

%
%

\section{Introduction}

Nowadays large data storage architectures such as Google FS~\cite{googlefs} or Amazon S3~\cite{s3} are built upon
networked distributed storage systems that spread data over several commodity storage servers. To ensure that data
survives disk failures, these storage systems must keep the original data together with some amount of redundancy. A
3-way replication has traditionally been used to that effect. However, today's systems such as Microsoft
Azure~\cite{azurestorage}, Hadoop FS~\cite{hdfsraid} or the new Google FS, are increasingly using redundancy schemes
based on erasure codes due to their capability to reduce storage
overheads~\cite{hadoopec,ecgoogle,diskreduce}. For example, in deployed systems erasure codes can
reliably store data with an overhead of 1.3x--1.5x the size of the original data~\cite{azurestorage,hdfsraid}.

One of the main problems in networked distributed storage systems is to replenish redundancy over time as storage nodes
fail. With replication and classical erasure codes (e.g. Reed-Solomon codes), repairing a missing piece of data entails
the same communication cost as transmitting a whole data object~\cite{codingvsrepl}, which causes a massive utilization
of network resources. Furthermore, a single node is responsible for the repair of a given object, which might create
network bottlenecks, slowing down the repair process and compromising data reliability.  This triggered a line of
research addressing the design of novel codes tailor-made for distributed storage which facilitate the repair process:
for example, Regenerating Codes (RGC)~\cite{Dimakis,Kermarrec,Schum} are designed to minimize the repair traffic,
while Self-Repairing Codes (SRC)~\cite{OD} aim at reducing the number of nodes contacted per repair.  However, these
novel code families have so far only been studied theoretically, in a simplistic setting where only one object is
stored. Storing multiple objects instead forces to take into account the repair of multiple failures for the same object
concurrently~\cite{corrfail}, which in turn might be affected by different chosen data placement strategies and
consequent contention of limited and shared network resources. This paper is a step to explore `\emph{How do these novel
codes, whose corresponding theories predict significant improvements on repairing objects in isolation, actually work
under realistic settings?}'.

Encoding an object to be stored over a network using an erasure code consists of splitting the object into $k$
fragments, that are transformed into $n>k$ redundant fragments, stored in distinct storage nodes in the system. The
transformation is such that the original object can be reconstructed from a subset of these redundant fragments. When a
storage node fails, it can be repaired by downloading some amount of data from a $d$-subset of live nodes ($d\leq n-1$).
RGC and SRC differ on how this $d$-subset is selected. RGC allow to select any $d$-subset out of the $n-1$ live nodes,
where $d$ is a relatively large value (the larger the $d$, the lower the repair traffic in RGC, and hence $d=n-1$ is
typically preferred), while SRC require much smaller values of $d$, usually as low as $d=2$. In that sense, these two
approaches, while addressing the same problem of better repairability in erasure coded storage systems, represent two
extreme design points in terms of the {\em repair degree} $d$, that is the number of nodes to be contacted per repair.
Both choices present their own pros and cons:
\begin{itemize}
\item In terms of choices of $d$-subsets, RGC are more flexible in that a lost fragment can be repaired from any
$d$-subset of live nodes, in contrast SRC require the use of specific $d$-subsets of nodes, although there are many such
possible subsets to choose from. 
\item The latency of the repair process can be severely affected if a single involved node is overloaded. Consequently,
a large repair degree $d$ makes RGC more vulnerable to overloaded nodes. Given the many possible choices of small
subsets with which a repair can be carried out, SRC avoid bottlenecks as long as they can locate one such subset with
unloaded live nodes.
\end{itemize}
Besides that, when multiple nodes fail simultaneously, both coding schemes can be used to repair each failure
independently from the others. The original design of SRC naturally preserves the good repairability properties for
multiple failures, but the repair performance of RGC degrades in this situation because live nodes are more prone to
provide data to several repair processes. However, a RGC variant called collaborative regenerating codes
(CRGC)~\cite{Kermarrec,Schum} have improved upon what can be achieved by allowing several repair processes to contact a
(possibly) different $d$-subset before exchanging data among each other.

\vspace{.3cm}
Despite the problems related with coding schemes, in real distributed storage systems nodes can be overloaded due to
unbalanced object assignments~\cite{mid2011}, or bad task scheduling \cite{mapredpref}.  In special we are interested on
these cases where some network endpoints at each node might become overloaded, identifying to main situations where this
can happen:
\begin{itemize}
\item Nodes in real systems store data from multiple objects. When one of this nodes fails an independent repair process
is triggered to repair each of the involved objects. These simultaneous repair processes might overload some of the live
nodes, creating network bottlenecks for repair processes.
\item Some data-intensive applications running in datacenters might require to send or receive large amounts of
data through the network (e.g., Map-Reduce tasks), causing temporally overloaded nodes.
\end{itemize}
When some part of nodes are overloaded due to any of these two reasons the limited number of repair choices of SRC can
make these codes more prone to experience slow repair times. However, RGC can download data from the first $d$ live nodes
with less network load, reducing the repair latency. This leads to the following central question: `\emph{Given some
node fault patterns and network load constraints, which code can carry out the repairs at what rate, and what are the
implications of the corresponding repair times on the system's resilience?}' The complete answers to these questions
are not obvious even for a single stored object, but eventually, studying a single object is not adequate, since
realistic environments require the storage of multiple objects, leading to further complexity~\cite{corrfail}.

This paper empirically evaluates the repair performance of RGC and SRC in a system storing and maintaining multiple data
objects. These two new code families were chosen since they represent the two possible extremes for optimizing the
repair process: (i) RGC aims at the minimal absolute repair communication (recall that the larger value of the repair
degree $d$ the less the repair traffic), while (ii) SRC minimize the number of live nodes needed to carry out a repair,
in fact, achieving $d=2$. The low repair degree in turn leads to not only significant reduction of repair traffic, but
also other benefits such as fast and parallel repairs. Additionally, we also contrast the results of these two novel
codes with traditional erasure codes.  Our analysis focuses on the required communication and repair speed, while
varying the code parameters, failure characteristics, data placement and network load. 

Our study leads to several
intuitive, as well as not so intuitive results. We confirm that both RGC and SRC reduce the maintenance communication
overhead significantly with respect to traditional erasure codes. Regarding data placement, it appears that the repair
process is significantly slowed down when data is placed in a clustered manner. A more interesting result concerns
repair speed: while RGC mostly consume less bandwidth than SRC, a pipelined variation of SRC achieve significantly
faster repair than all codes under all the settings that we have studied. We also observe that since bandwidth is an
ephemeral resource, the more important thing is to utilize it in a balanced manner over time, and the very low value of
\emph{d} in SRC facilitate the same, to carry out fast and parallel repairs - even when multiple objects, as well as
multiple faults are considered. Finally, regarding the performance of repairs on overloaded networks we identify that
repairs finalize significantly faster when the same overall load is caused by short and frequent ``busy'' node periods
than with long and sporadic ones.

The rest of the paper is organized as follows. In Section~\ref{sec:relwork} we first present some background on erasure
codes and related works proposing different constructions or analysis of codes for distributed storage systems. In
Section \ref{sec:setting} we provide essential information on the codes we study, followed by a description of our
simulation methodology, including how we model the various properties of the codes as well as the networked storage
environment. We present our findings in Section \ref{sec:evaluation}. Finally, in Section \ref{sec:conclude} we conclude
by discussing the practical implications of the results.

%
%
\section{Background and Related Work}
\label{sec:relwork}

Erasure codes have been largely studied in the storage literature as a mechanism to provide high data reliability and
reduce the storage overhead required with respect to simple data replication. Given a data object of size $B$, an
($n,k$) erasure code splits this object into $k$ smaller fragments each of size $B/k$. These $k$ fragments are then
mapped to a set of $n$ redundant fragments, $n>k$, to be stored into $n$ different nodes. If the code is a maximum
distance separable (MDS) code, the stored object can be reconstructed from any $k$-subset of redundant fragments. For
example, 3-way replication is a (3,1) MDS erasure code that maps the object to three copies of itself.

One of the main uses of erasure codes is to protect multiple disk failures in RAID-6\linebreak  disk
configurations~\cite{liber8tion}, where the $n$ redundant fragments are distributed into small sets of disks --usually
from four to eight disks. Since the value of $n$ is small, the code can be implemented as a flat-XOR code: all redundant
fragments are obtained by xoring some of the $k$ original fragments~\cite{liber8tion}. Furthermore, it was shown
in~\cite{Khan11} that these flat-XOR codes can repair redundant fragments without needing to reconstruct the entire
original object, by reading only $d$ live fragments, where $d<k$.

Erasure codes are also used in networked distributed storage systems to reduce the storage overhead down to 1.3x the
size of the original object, $n/k\simeq 1.3$~\cite{azurestorage,hdfsraid}. In these systems the $n$ redundant fragments
are stored across different commodity storage servers, which are more prone to disk failures, power outages, network
disconnections and software errors. Achieving a low storage overhead in these environments while guaranteeing a high
data availability requires to spread data over larger sets of storage nodes. Since designing optimal flat-XOR codes for
large $n$ values, and finding efficient repairs ($d<k$) are two NP-hard problems~\cite{flatxor}, existing solutions use
traditional erasure codes like Reed-Solomon codes, i.e., codes designed for communication over noisy channels. They have
already been extensively studied in the context of networked storage (e.g.~\cite{OceanStore}), and require more complex
finite-field operations. Their main drawback is that repairing a missing fragment entails contacting $d=k$ storage
nodes, and the same communication cost as transmitting a whole data object~\cite{codingvsrepl}.

Recently, some novel erasure codes have been designed to reduce this repair cost. Regenerating Codes
(RGC)~\cite{Dimakis} and Collaborative RGC (CRGC)~\cite{Kermarrec,Schum} can be seen as a composition of an MDS erasure
code and a network code~\cite{netcoding}, aiming at the minimal communication to repair one failure at a time for RGC, and several node
failures in parallel for CRGC by enabling collaboration among repairing nodes. As compared to traditional erasure codes
like Reed-Solomon, RGC and CRGC allow to reduce the repair communication at the cost of increasing the number of
contacted nodes, $d>k$. Some variants of RGC like Minimum Bandwidth RGC~\cite{Dimakis} allow to further reduce the
repair communication at the additional cost of increasing the size of the redundant fragment (larger than $B/k$), making
the codes less suitable for environments where minimizing the storage footprint is a priority. Furthermore, the recent
Simple Regenerating Codes~\cite{simplerc} allow to easily combine $t=2$ classical MDS erasure codes to reduce the number
of nodes contacted during repairs to $d=4$. Unfortunately, this approach increases the overall storage overhead by 50\%
compared to RGC or CRGC. In general the overhead can be reduced to $100t^{-1}\%$ by increasing the number of contacted
nodes to $d=2t$, but losing the potential to repair redundant fragments efficiently~\cite{simplerc2}.

Finally, Self-Repairing Codes (SRC)~\cite{OD} are a family of non-MDS erasure codes which like RGC and CRGC aim at
minimizing the repair traffic and storage overhead, though this is attained by drastically reducing the number of live
nodes contacted for a repair. To be specific, in SRC only $d=2$ live nodes need to be contacted for a repair, allowing
to repair up to $(n-1)/2$ simultaneous faults. In this paper we will compare the repair performance of these two
families of coding techniques (RGC/CRGC and SRC) to classical Erasure Codes (EC).

%
%

\section{Theoretical and Experimental Settings}
\label{sec:setting}

In this section we will present the main theoretical features of the codes which are the subject of this study and we
will describe the simulator framework that that we will use in Section~\ref{sec:evaluation} to evaluate the repair
performance of these codes.

\subsection{Novel Coding Techniques}

By an {\bf erasure code} (EC), we mean a map that encodes $k$ fragments into $n$, with the property that any choice of
$k$ encoded fragments is enough to recover the encoded object --i.e., they have the MDS property. Each node stores an
amount of data per object equal to $B/k$, which is the minimal amount possible to guarantee that objects can be
reconstructed by retrieving $k$ encoded fragments out of the total $n$. In order to repair one failure, a repair
process, also called newcomer, downloads $k$ encoded fragments, from which it recovers the object, and can thus
recompute the missing encoded fragment. This is costly both in terms of download data and computation, though it becomes
interesting in the case of \emph{lazy repair}, where the system waits for $f$ ($f\leq n-k$) failures to accumulate
before triggering the repairs.  Such a repair procedure, where one node reconstructs the encoded fragments and then
distributes them to other nodes has an average communication cost per failure (for one object and normalized by $B/k$) of
\begin{equation}
\label{eq:gammaec} \gamma_{EC} = \frac{k+(f-1)}{f}.
\end{equation}

\vspace{.5cm}
{\bf Regenerating Codes} (RGC) can be seen as erasure codes in that object reconstruction is similarly done by
contacting any choice of $k$ nodes (they have the MDS property). Repair is however done differently as we now explain by
considering Collaborative RGC (CRGC)~\cite{Kermarrec,Schum}, which include RGC as a special case. CRGC allow newcomers
to collaboratively repair $f$ failed fragments. In CRGC, repair is done in two steps: a download phase where the $f$
repair processes download data from $d$ live fragments, $d\geq k$, and a collaborative phase, where all the $f$ nodes
exchange data among themselves. When $f=1$, CRGC are exactly RGC, however, in the event of multiple failures, it is more
favorable to use CRGC than RGC which can only repair one failure at a time sequentially. The total amount of
normalized repair traffic per failure of one object, when each node stores an amount of data $B/k$, is
\begin{equation}\label{eq:gammacrgc}
\gamma_{CRGC}= \frac{d+f-1}{d-k+f},
\end{equation}
which was derived analytically assuming a single stored object~\cite{Kermarrec}. We note that when $k=d$,
$\gamma_{CRGC}=\gamma_{EC}$, thus EC can be considered as a special case of CRGC.

\vspace{.5cm}
{\bf Self-Repairing Codes} (SRC) minimize the number of live nodes to be contacted for repair. In fact, SRC enables the
repair of a single failure by contacting only 2 nodes, while $f$ failures can be repaired by contacting only 2 nodes per
repair for up to $f \leq (n-1)/2$ failures. More failures can be tolerated, but without the guarantee that contacting
only 2 nodes will work. SRC thus cannot have the MDS property: take one fragment that can be obtained from two others,
an object cannot be retrieved from $k$ fragments including these 3 fragments. Still assuming $B/k$ amount of data per
object per node, the normalized repair communication per failure is

\begin{equation}\label{eq:gammasrc}
\gamma_{SRC}=\frac{2f}{f}=2.
\end{equation}

It is important to note that $\gamma_{EC}$, $\gamma_{CRGC}$ and $\gamma_{SRC}$ are theoretical bounds subject
to the existence of explicit code constructions satisfying them. In this paper we focus on the potential repair performance
of the different codes independently of the currently known code constructions.

\subsection{Pipelined Repairs}

Besides considering the regular repair procedure of RGC/CRGC and SRC, we introduce \emph{pipelining repairs} for SRC
(SRCp), adapted from~\cite{Li}, originally conceived for RGC with heterogeneous link capacities. Concretely, the repair
process for SRCp does not download data directly from 2 nodes, but asks the first node to download data from the second.
Then this first node encodes the received data byte-by-byte with the data it stores and forwards it to the node running
the repair process. Thus, the repair can finish within the time to transmit only one fragment plus a small overhead,
negligible for large objects, due to the time required to encode and transmit the first fragment bytes.  We do not
consider pipelined RGC since they use additional storage at auxiliary `apprentice' nodes, hence are not strictly
comparable, and their repair time is lower bounded by that of SRCp, though at a significantly higher implementation
complexity.

\subsection{Simulator Set Up}

To evaluate the behavior of different codes, we use a discrete time simulator where, at each time step, all nodes can
send and receive (full-duplex network with symmetric bandwidth) a data packet of a fixed length. Outgoing packets from
each node are queued until the destination node is available to receive it.  Once a node receives data from an outgoing
queue of some other node it will then refuse further requests during the same time round. Which request is accepted, and
which others are declined is decided randomly without bias. Rejected nodes retry to find any other appropriate node to
send their outgoing packets if that is possible. This simulation model allows to emulate network congestion on those
nodes with more incoming/outgoing traffic. For the sake of simplicity but without loss generalization, we set the packet
length to $\beta$, which is the minimum amount of data that nodes transmit during repairs. We set the discrete step
duration $\tau$ to $\tau=\beta/(\omega\cdot\text{\em eff})$, where $\omega$ is the upload/download bandwidth, set to
$\omega=1$Gbps, and \emph{eff}, \emph{eff}$\in(0,1]$, is the network efficiency, set to \emph{eff}=0.8 (as a default
value). This $\emph{eff}$ parameter allows us to emulate network overheads such as packet headers or retransmissions.

\vspace{.3cm}
Besides the network load caused by the repair traffic we also aim to simulate load due to real data-intensive processes
like Map-Reduce tasks. To emulate these data-intensive processes we assume that nodes can have overloaded periods where
they cannot send or receive repair data. Repair processes needing data from these overloaded nodes will have to wait or
find other suitable nodes, which might lengthen repair times and compromise data reliability.  We will assume that
overloaded periods at each node have Poisson arrivals and that the durations of these periods are exponentially
distributed. If $\lambda_a$ and $\lambda_d$ represent respectively the arrival rate and the duration rate of these
overloaded periods, then average number of overloaded nodes at any time is $N\cdot\lambda_a/\lambda_d$, where $N$ is the
total number of nodes in the system.  Although this model does not capture all peculiarities of a real system running
data-intensive applications it gives us a simple way to analyze how different congested scenarios can affect on the
repair performance of different codes.

\vspace{.3cm}
Finally, in our simulated environment the different repair processes are executed as follows: For RGC/CRGC, the repair
process downloads $d$ fragments from the first $d$ nodes out of the live $n-f$ that have a free uploading slot ($f$ is
the number of failures at a given time). For SRC and SRCp, it has a list with all the possible pairs of nodes available
to repair each lost fragment. In the case of SRCp, it uses the first pair of nodes that are simultaneously available to
upload data. This repair takes then $\tau$ seconds, plus the time (number of discrete time intervals) it had to wait
before a suitable pair of blocks became simultaneously available. For SRC the repair process needs two nodes to upload
their fragments in two different time steps. Due to the limited pairs available for each repair, the repair pair
selection can have a significant impact on the repair time. Analyzing different SRC repair schedules is out of the scope
of this paper and we choose a simple strategy: we randomly select the pair of nodes used for each repair. In
Section~\ref{s:multiple_failures} we will show that this simplistic policy can have detrimental effects when there are
correlated failures. More sophisticated scheduling mechanisms for SRC would likely yield improved repair performance,
but exploration of such scheduling mechanisms is out of the scope of this work. In that sense, the results provide a
pessimistic baseline of how SRC based repairs perform, leaving room for further improvements.

%
%
\begin{figure*}
  \centering
  \subfloat[(7,4) code]{\label{fig:rel1}\hspace{-2mm}\includegraphics[scale=.85]{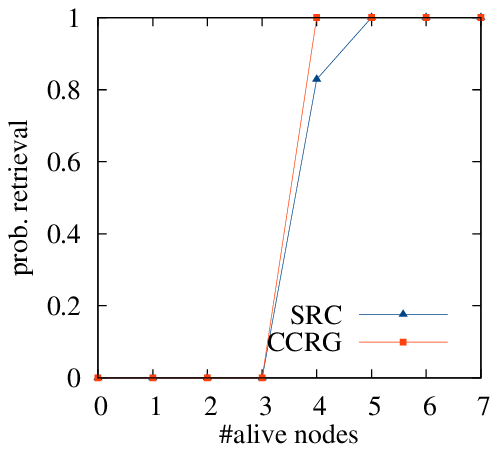}}
  \subfloat[(7,3) code]{\label{fig:rel2}\hspace{-7mm}\includegraphics[scale=.85]{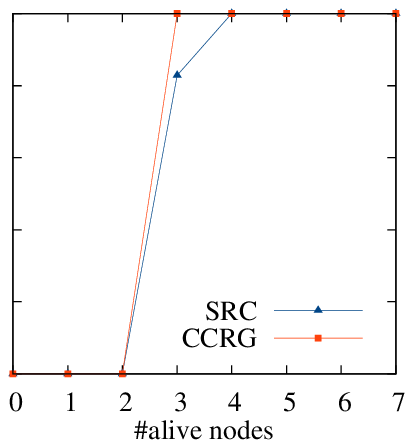}}
  \subfloat[(15,5) code]{\label{fig:rel3}\hspace{-7mm}\includegraphics[scale=.85]{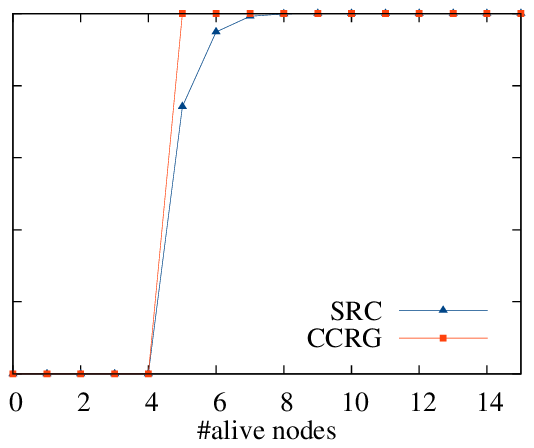}}
  \caption{Probability to retrieve a stored object for different number of live nodes and for different code
  parameters.}
  \label{f:prob}
\end{figure*}

\section{Evaluation}
\label{sec:evaluation}

In the evaluation of the different codes we distinguish three different scenarios:
\begin{enumerate}
\item Only one node fails ($f=1$) and nodes do not have overloaded periods. 
\item Only one node fails ($f=1$) and nodes have overloaded periods.
\item There are correlated node failures and multiple encoded blocks for the same object may
be missing.
\end{enumerate}
In scenarios 1 and 2 a single fragment is lost for every object stored on the failed node.
For each scenario we also evaluate three different $(n,k)$ code parameters, namely (7,4), (7,3) and (15,5), respectively achieving
the storage overheads of 1.75, $2.\bar 3$ and 3. In Figure~\ref{f:prob} we depict the static resilience or probability
of being able to recover the stored object in the presence of node failures of the three different code parameters.
These results are obtained by enumerating all possible node failure combinations. We can appreciate how the static
resilience of the non-MDS SRC is comparable to that of CCRG.

\begin{figure*}
  \centering
  \mbox{\hspace{-3cm}\includegraphics[scale=0.7]{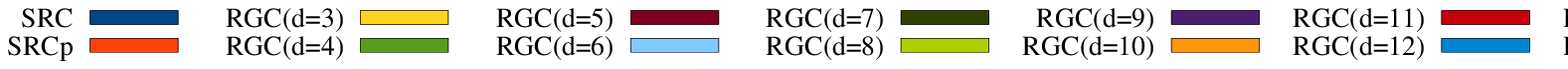}}\vspace{-.4cm}
  \subfloat[Avg. repair time for (7,4).]{\label{fig:granul:t4}\includegraphics[scale=1]{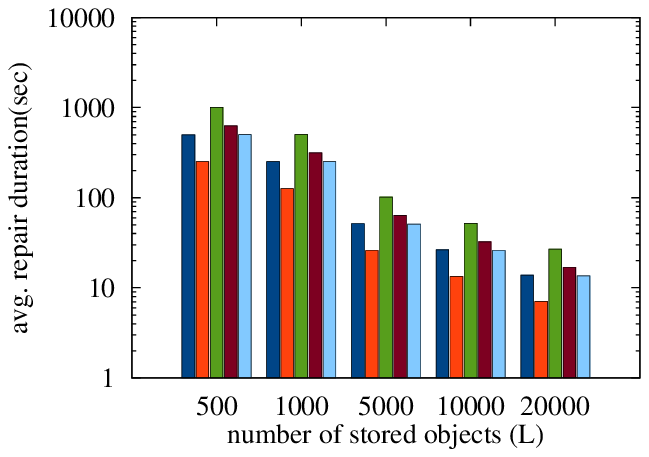}}
  \subfloat[Avg. traffic for (7,4).]{\label{fig:granul:bw4}\includegraphics[scale=1]{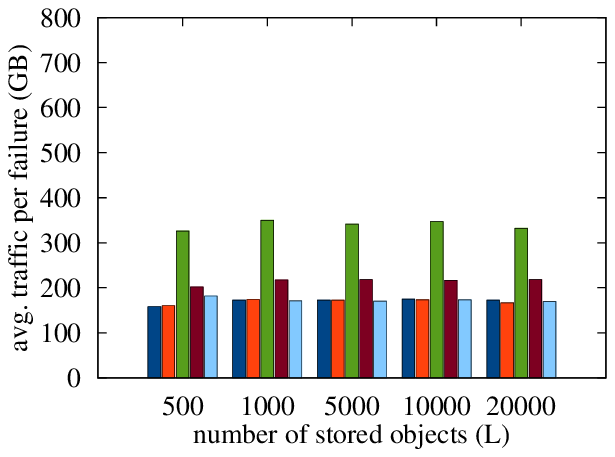}}
  \quad
  \subfloat[Avg. repair time for (7,3).]{\label{fig:granul:t3}\includegraphics[scale=1]{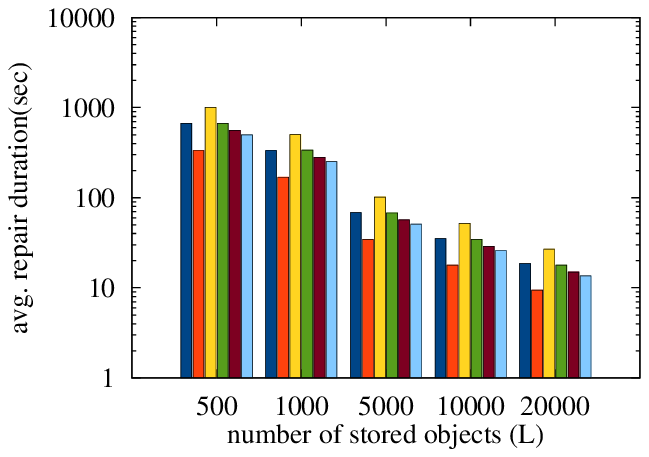}}
  \subfloat[Avg. traffic for (7,3).]{\label{fig:granul:bw3}\includegraphics[scale=1]{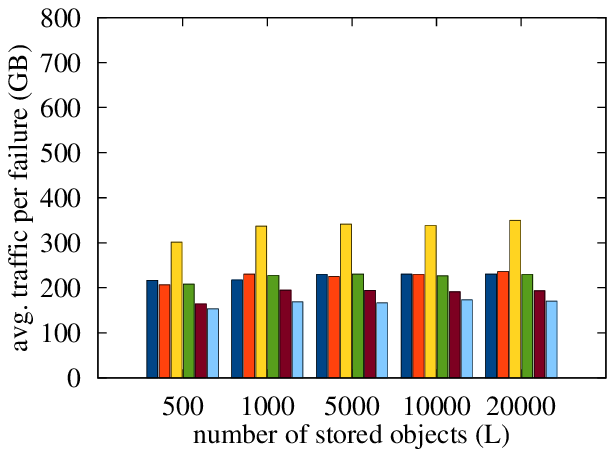}}
  \quad
  \subfloat[Avg. repair time for (15,5).]{\label{fig:granul:t5}\includegraphics[scale=1]{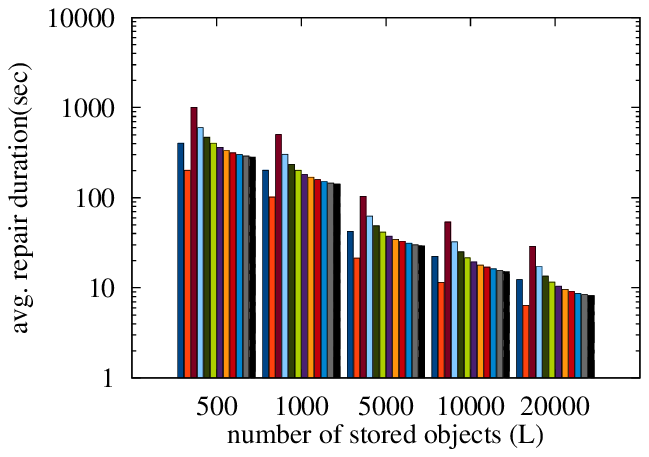}}
  \subfloat[Avg. traffic for (15,5).]{\label{fig:granul:bw5}\includegraphics[scale=1]{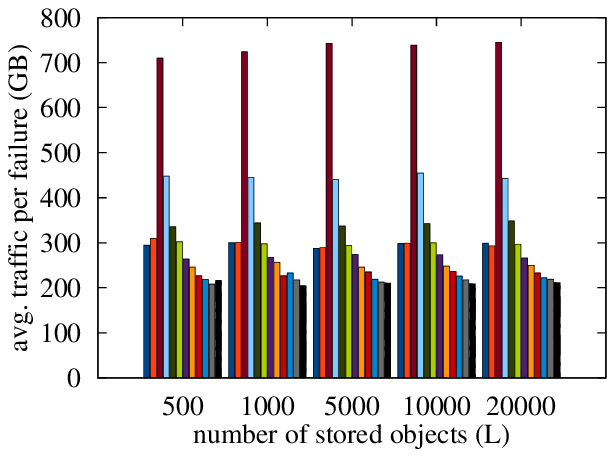}}
  \caption{Analysis of one single node failure: performance of different codes is shown as a function of the number $L$ of
  stored objects. The overall amount of data stored across $N=1,000$ nodes in the system is $B\cdot L=50$TB.}
  \label{fig:granul}
\end{figure*}

\subsection{Single Node Failure Evaluation}
\label{s:single_failure}

To analyze the effects of a single node failure we randomly select one storage node and delete all its stored data. Then
we start repair processes to regenerate all the missing fragments, measuring the repair times and traffic consumed by
each of them. Our plotted results are the average results from 1,000 independent experiments. The single node failure
analysis is done for Regenerating Codes (RGC), Self-Repairing Codes (SRC) as well as for its pipelined version (SRCp).
Recall that (i) all nodes store the same amount of data $B/k$ per object, (ii) that for one single failure, CRGC reduces
to RGC, and (iii) that Erasure codes (EC) can be considered as a special case of CRGC code with $d=k$.

\paragraph{Evaluating Data Granularity:}

To measure the impact of different data granularities or data partition sizes on the system performance, we assume a
system storing a total amount of $B\cdot L=50$TB, where $B$ is the object size and $L$ the number of stored objects.
These 50TB correspond to the size of the stored data, without redundancy. Storing this amount of data using an ($n,k$)
code requires an aggregated node capacity of $n\cdot B\cdot L/k$.  We evaluate different data granularities by running
simulations for different values of $L$, where the $n$ redundant fragments are randomly stored into $N=1,000$ nodes.

\vspace{.3cm}
In Figure~\ref{fig:granul}, we depict the results for the granularity experiment using a random data placement. In
Figures \ref{fig:granul:bw4}, \ref{fig:granul:bw3} and~\ref{fig:granul:bw5} we show the \emph{average overall traffic}
required to repair a failed node as a function of the granularity --i.e., the number of objects $L$. Despite some small
variations due to the random fragment placement and the averaging of 1,000 experiments, we see how the experimental
results fit the analytical predictions defined in (\ref{eq:gammaec}), (\ref{eq:gammacrgc}) and (\ref{eq:gammasrc}): from
(\ref{eq:gammacrgc}) $\gamma_{CRGC}= \frac{d}{d-k+1}$, thus increasing the repair degree $d$ in RGC allows to reduce the
overall network traffic. SRC/SRCp achieve lower repair traffic than RGC only when $d<2k-2$, since from
(\ref{eq:gammasrc}) $\gamma_{SRC}=2$. We can also appreciate how the (15,5) code, which has the largest storage
overhead, requires more traffic per failure, since more data needs to be repaired per failure.

\begin{figure*}
    \centering
  \mbox{\hspace{-3cm}\includegraphics[scale=0.7]{figs/legend1.eps}}\vspace{-.4cm}
  \subfloat[Avg. repair time for (7,4).]{\label{fig:plac:t4}\includegraphics[scale=1]{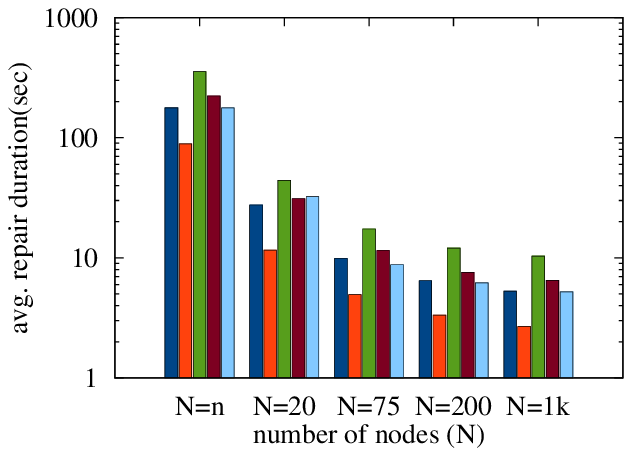}}
  \subfloat[Avg. traffic for (7,4).]{\label{fig:plac:bw4}\includegraphics[scale=1]{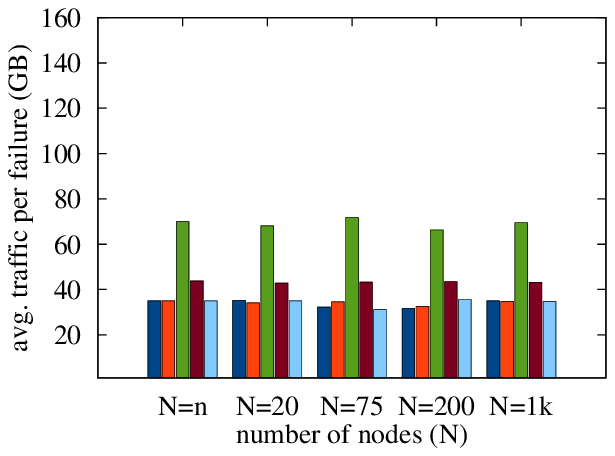}}
  \quad
  \subfloat[Avg. repair time for (7,3).]{\label{fig:plac:t3}\includegraphics[scale=1]{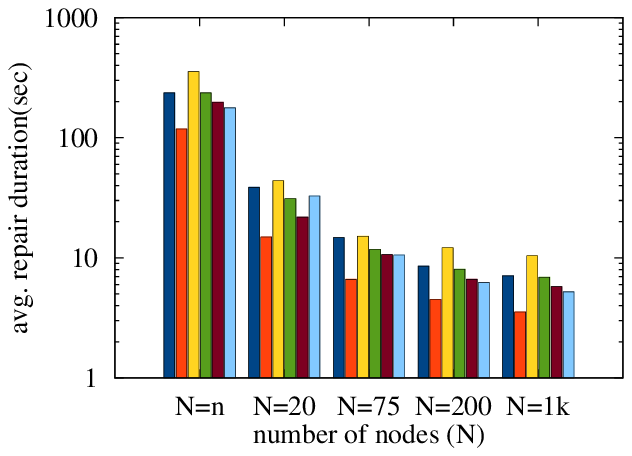}}
  \subfloat[Avg. traffic for (7,3).]{\label{fig:plac:bw3}\includegraphics[scale=1]{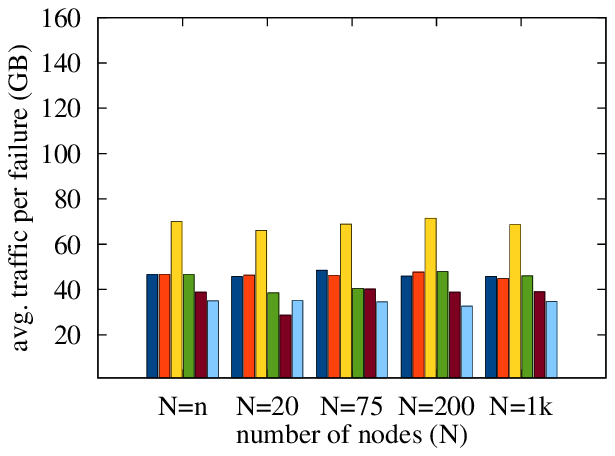}}
  \quad
  \subfloat[Avg. repair time for (15,5).]{\label{fig:plac:t5}\includegraphics[scale=1]{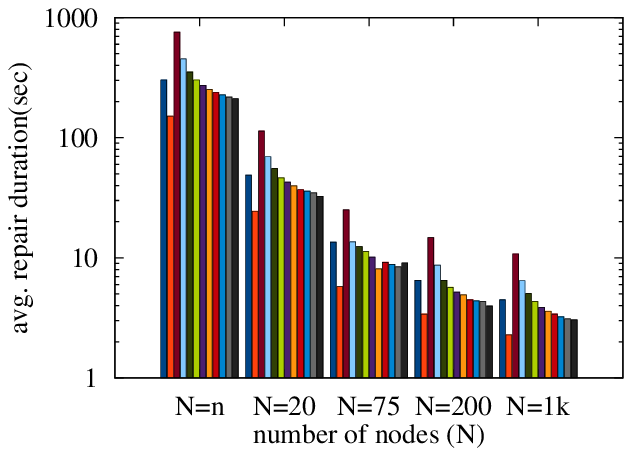}}
  \subfloat[Avg. traffic for (15,5).]{\label{fig:plac:bw5}\includegraphics[scale=1]{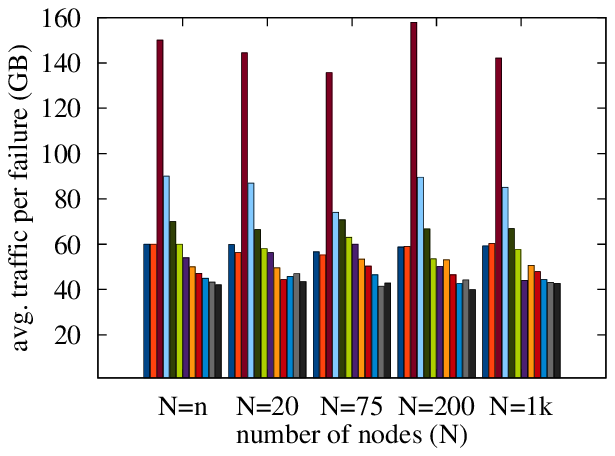}}
  \caption{Analysis of one single node failure: performance of different codes is shown for different data clusterings. The size of the objects is $B=1$GB, and we store an amount of objects proportional to the cluster size, $L=10\cdot N$.}
  \label{fig:plac}
\end{figure*}

\vspace{.3cm}
Figures~\ref{fig:granul:t4}, \ref{fig:granul:t3} and~\ref{fig:granul:t5} show in logarithmic scale the \emph{average
fragment repair times} for a single node failure. As we can see, SRCp exactly halve the repair times of SRC. This
implies that for single failures, SRC achieve a good repair performance by using any random pair of fragments.  It is
also interesting to see how even for those RGC configurations that require less repair traffic than SRCp, the repair
time for SRCp is significantly shorter than for RGC. Finally, note that from $L=500$ to $L=20,000$, $B$ is reduced by a
factor of 40, which is roughly the same improvement that we measure on repair times when we switch from $L=500$ to
$L=20,000$. We thus conclude that \emph{data granularity has no significant impact on the performance of the analyzed
codes}.

\paragraph{Evaluating Data Placement Strategies:}

Networked distributed storage systems also need to deal with data placement: \emph{`How to assign redundant fragments to
nodes to maximize system performance and data reliability?'}  To measure the impact of different data placement
strategies, we use the concept of data clustering~\cite{Venkatesan10_replica}. We divide the full set of storage nodes
into disjoint subsets of nodes called clusters. Each of these clusters is an independent storage system with $N$ nodes,
which stores all the fragments for a given object. We assume that fragments within the cluster are randomly distributed.
For the smallest value of $N$, $N=n$, the data placement becomes a \emph{full clustered} placement where all nodes in
the cluster store fragments of the same set of objects.
\begin{figure*}[t]
  \centering
  \subfloat[(7,4) code.]{\hspace{-.4cm}\label{fig:bwcharsingle:74}\includegraphics[scale=1]{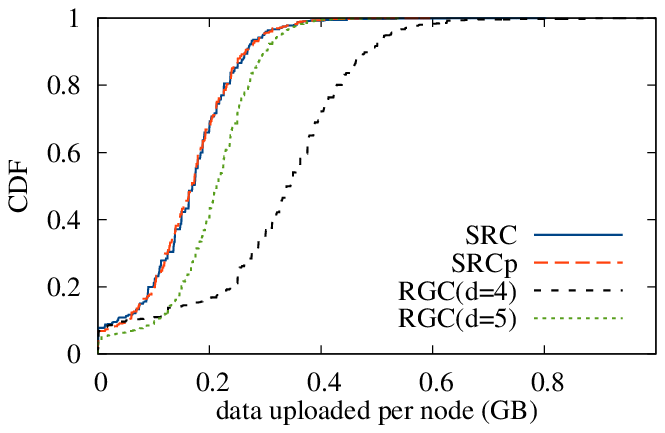}}
  \subfloat[(7,3) code.]{\label{fig:bwcharsingle:73}\hspace{-.8cm}\includegraphics[scale=1]{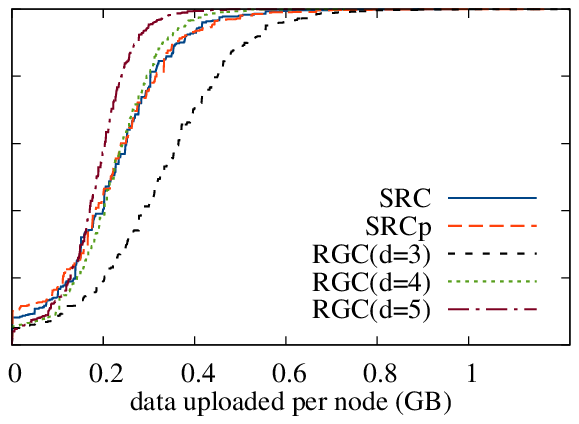}}
  \quad
  \subfloat[(15,5) code.]{\label{fig:bwcahrsingle:155}\hspace{-.8cm}\includegraphics[scale=1]{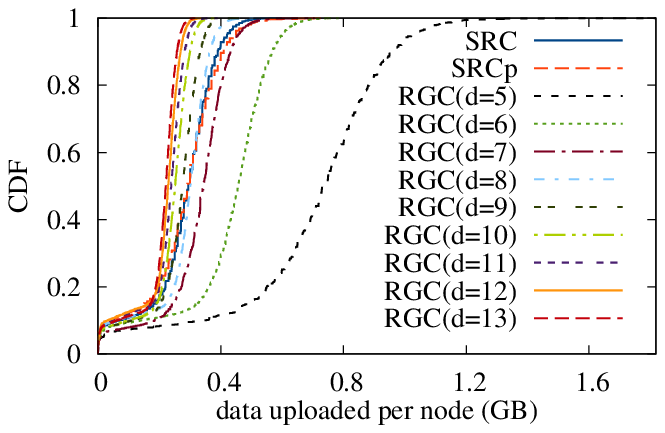}}
  \caption{CDF of the amount of data each node uploads in order to repair a single node failure. Results are obtained
  for a configuration with $L=10,000$, $N=1,000$ and $B=1$GB.}
  \label{fig:bwcharsingle}
\end{figure*}

\vspace{.3cm}
In Figure~\ref{fig:plac}, we depict the results for different placement strategies. In Figures~\ref{fig:plac:bw4},
\ref{fig:plac:bw3} and~\ref{fig:plac:bw5} we observe no differences in the \emph{average overall traffic per failed node}
for different placements. However, in Figures~\ref{fig:plac:t4}, \ref{fig:plac:t3} and~\ref{fig:plac:t5} we see how the
\emph{average fragment repair time} increases exponentially with data clustering since the same network resources are
needed for all the repairs. Also for any given degree of clustering, SRCp consistently and significantly outperforms all
other codes.

\paragraph{Bandwidth Usage Characterization:}

To conclude the evaluation of single failures, Figure~\ref{fig:bwcharsingle} illustrates the CDF for the amount of data
each node uploads per failure. Steeper curves represent better load balancing among nodes. For RGC we can see how large
$d$ values achieve better traffic balancing among nodes. In contrast, despite using only 2 nodes per repair, SRC/SRCp
achieve a good network traffic balancing, always better than the RGC (7,4) code, and better than RGC when $d\leq4$ and
$d\leq7$ respectively for the (7,3) and (15,5) codes. More balanced usage of bandwidth across all nodes translates to
fewer contentions for the same resources, hence faster repairs.

\subsection{Repair Performance in Loaded Networks}
\label{s:busy}

We consider four different scenarios with different average number of temporally overloaded nodes, namely 10\%, 20\%,
50\% and 80\% of nodes. For each of these scenarios we consider two different ways of achieving this percentage of
overloaded nodes, (i) one where nodes have short overloaded periods but become overloaded at a high rate, and (ii) a second one
where nodes have long overloaded periods, but become overloaded at a lower rate. Table~\ref{t:params} depicts the
different parameters used in each case.

\begin{table}[t]
\centering
\begin{tabular}{p{3.5cm}cccccccc} \\ \toprule
 & \multicolumn{2}{c}{10\%} & \multicolumn{2}{c}{20\%} & \multicolumn{2}{c}{50\%} & \multicolumn{2}{c}{80\%}\\
 & ~~short~~ & ~~long~~ & ~~short~~ & ~~long~~ & ~~short~~ & ~~long~~ & ~~short~~ & ~~long~~ \\ \toprule
Avg. no. of nodes switching to ``overloaded'' state at each step. & 25 & 4 & 50 & 4 & 125 & 4 & 200 & 4 \\ \midrule
Avg. duration of the \\``overloaded'' state. & 4 & 25 & 4 & 50 & 4 & 125 & 4 & 200 \\ \bottomrule
\end{tabular}
\vspace{2mm}
\caption{Different parameter values to simulate different percentage of overloaded nodes when there are $N=1000$ nodes
in the system.}
\label{t:params}
\end{table}

\begin{figure*}
  \centering
  \mbox{\hspace{-3cm}\includegraphics[scale=0.7]{figs/legend1.eps}}\vspace{-.4cm}
  \subfloat[(7,4) code for long overloaded periods.]{\label{fig:busy:t4l}\includegraphics[scale=1]{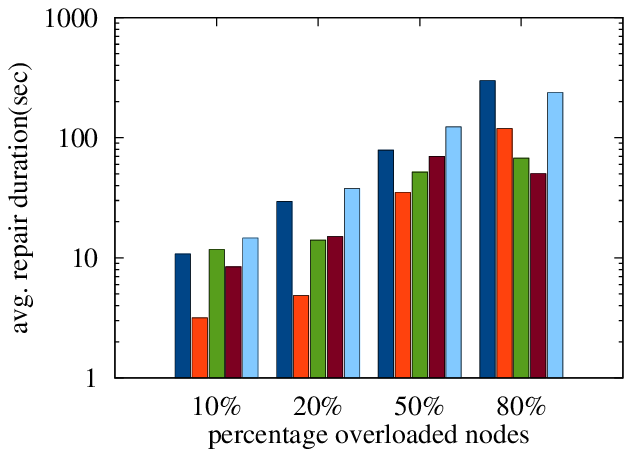}}
  \subfloat[(7,4) code for short overloaded periods.]{\label{fig:busy:t4}\includegraphics[scale=1]{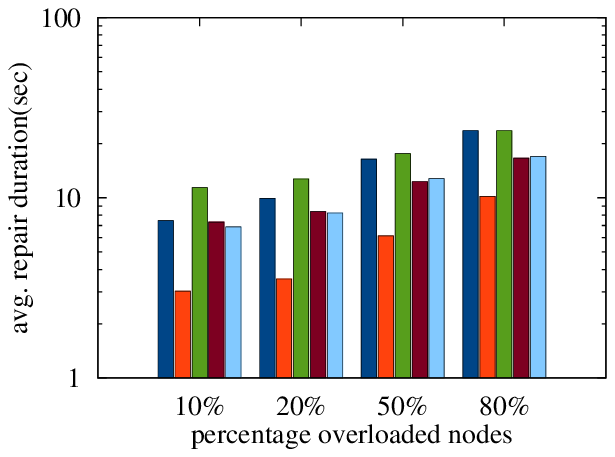}}
  \quad
  \subfloat[(7,3) code for long overloaded periods.]{\label{fig:busy:t3l}\includegraphics[scale=1]{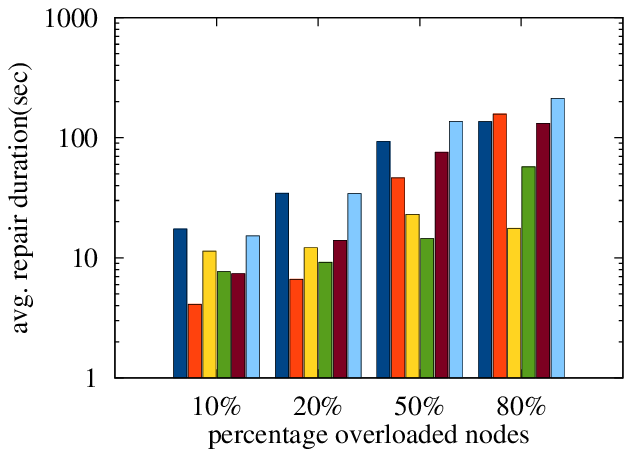}}
  \subfloat[(7,3) code for short overloaded periods.]{\label{fig:busy:t3}\includegraphics[scale=1]{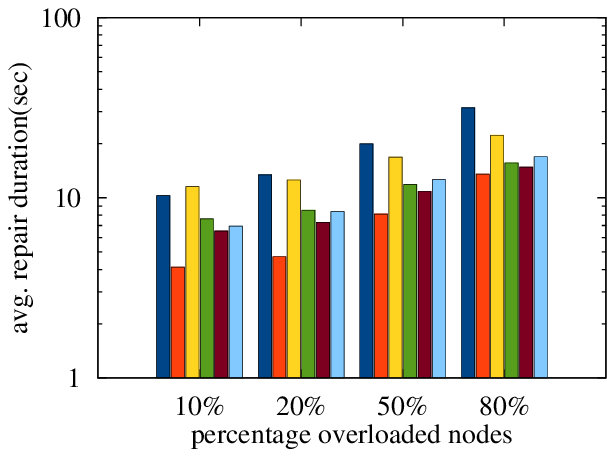}}
  \quad
  \subfloat[(15,5) code for long overloaded periods.]{\label{fig:busy:t5l}\includegraphics[scale=1]{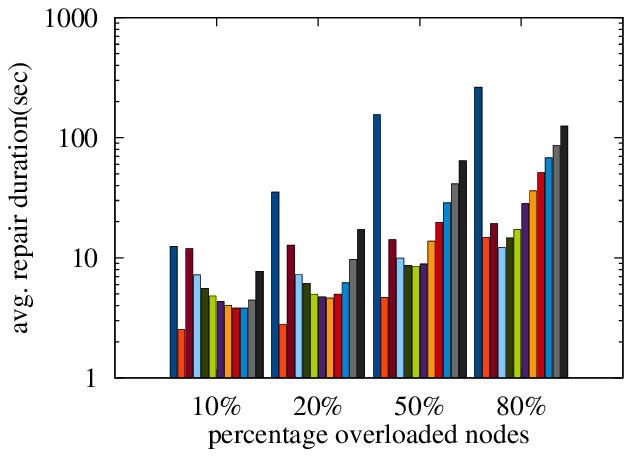}}
  \subfloat[(15,5) code for short overloaded periods.]{\label{fig:busy:t5}\includegraphics[scale=1]{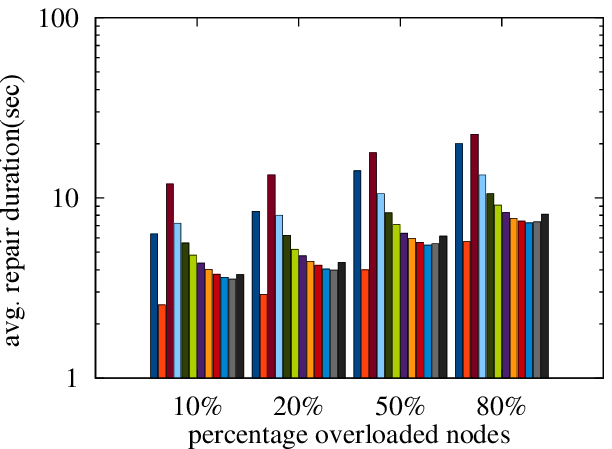}}
  \caption{Average repair times for one single node failure when a fraction of the nodes are temporally overloaded and cannot
  receive/send data. The size of the objects is $B=1$GB, and we store and the number of nodes is $N=1000$.}
  \label{fig:busy}
\end{figure*}

\vspace{.3cm}
In Figure~\ref{fig:busy} we show the average repair time for the different codes under different percentages of
overloaded nodes. Figures~\ref{fig:busy:t4l}, \ref{fig:busy:t3l} and~\ref{fig:busy:t5l} show the average repair times
for nodes with long overloaded periods. In general we can appreciate how for all code parameters repair times become
longer as more nodes are simultaneously overloaded. However, it is interesting to see how unlike in non-overloaded networks
(Figures~\ref{fig:granul} and~\ref{fig:plac}), RGC need longer repair times when the repair degree $d$ increases.
Although large $d$ values reduce the repair traffic, when part of the nodes are overloaded it becomes more difficult to
contact with $d$ nodes, and the repair process might need to wait. Similarly, Figures~\ref{fig:busy:t4},
\ref{fig:busy:t3} and~\ref{fig:busy:t5} depict the average repair times when nodes have short overloaded periods. In
this case increasing the value of the repair degree $d$ does not have the detrimental effect observed in the previous
set of figures and \emph{repair times are one order of magnitude shorter} than for long overloaded periods.

\begin{figure*}
  \centering
  \mbox{\hspace{-3cm}\includegraphics[scale=0.7]{figs/legend1.eps}}\vspace{-.4cm}
  \subfloat[(7,4) code for long overloaded periods.]{\label{fig:busy:bw4l}\includegraphics[scale=1]{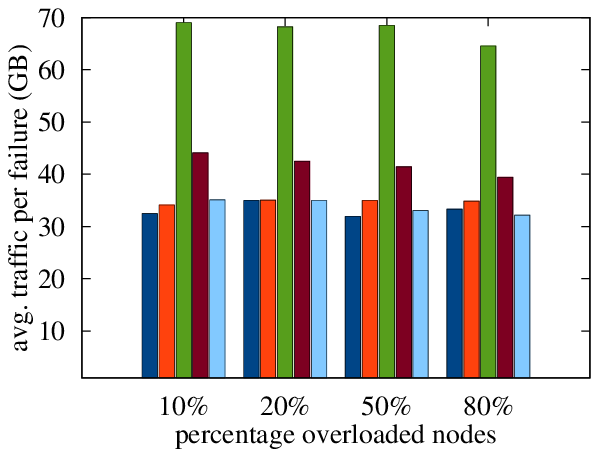}}
  \subfloat[(7,4) code for short overloaded periods.]{\label{fig:busy:bw4}\includegraphics[scale=1]{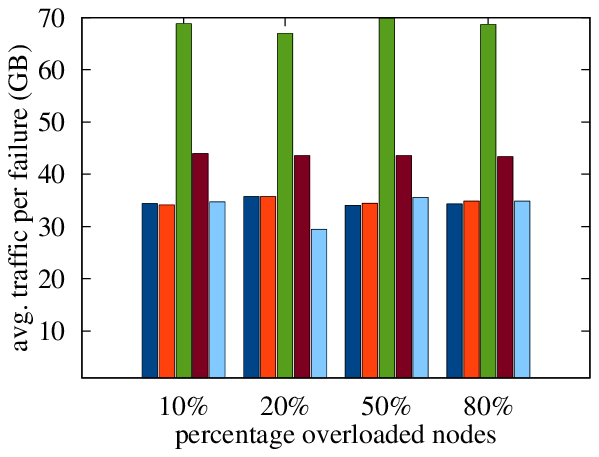}}
  \quad
  \subfloat[(7,3) code for long overloaded periods.]{\label{fig:busy:bw3l}\includegraphics[scale=1]{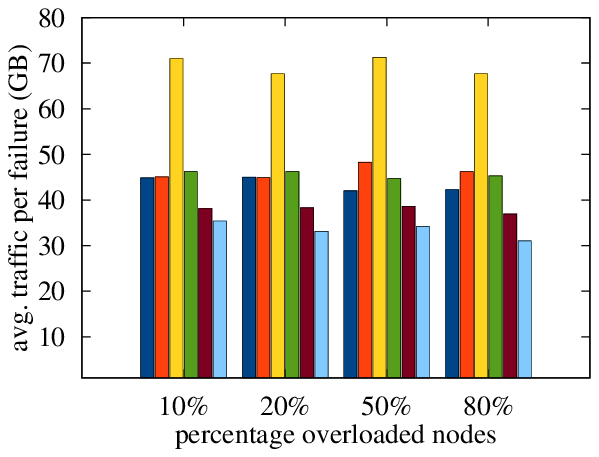}}
  \subfloat[(7,3) code for short overloaded periods.]{\label{fig:busy:bw3}\includegraphics[scale=1]{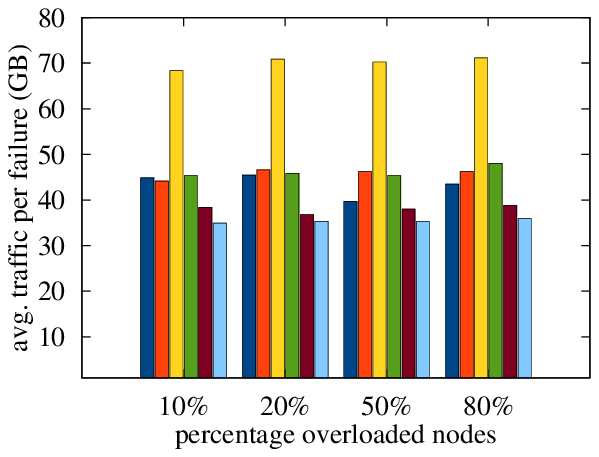}}
  \quad
  \subfloat[(15,5) code for long overloaded periods.]{\label{fig:busy:bw5l}\includegraphics[scale=1]{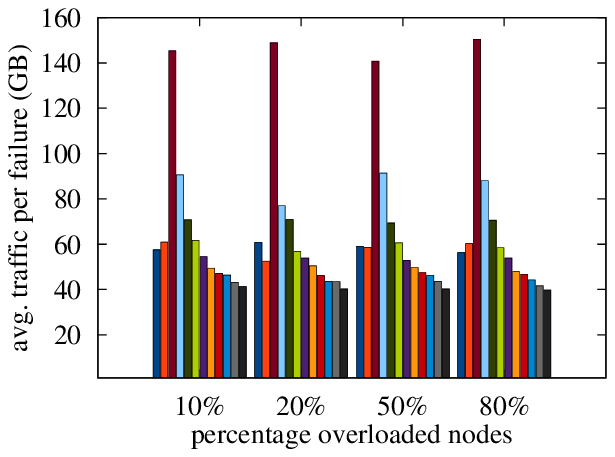}}
  \subfloat[(15,5) code for short overloaded periods.]{\label{fig:busy:bw5}\includegraphics[scale=1]{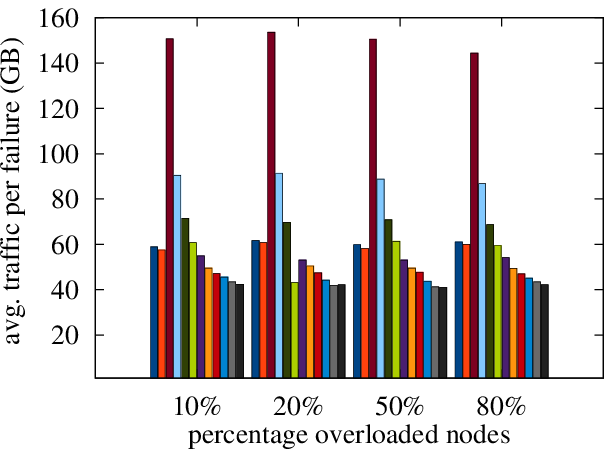}}
  \caption{Average traffic per failure for one single node failure when a fraction of the nodes are temporally overloaded and cannot
  receive/send data. The size of the objects is $B=1$GB, and we store and the number of nodes is $N=1000$.}
  \label{fig:busy2}
\end{figure*}

\vspace{.3cm}
Regarding the network traffic required to repair lost fragments in loaded systems (Figure~\ref{fig:busy2}) we cannot
observe noticeable differences. In both cases, for short and long overloaded periods the traffic required to repair a
failed node is the same than for non-overloaded systems.

\subsection{Multiple Node Failures}
\label{s:multiple_failures}

\begin{table}[t]
\centering
\begin{tabular}{lcccccc} \toprule
{\bf $~~~~~~~~~~~~~~~\Theta=$} & ~~~~~~{\bf 0.05}~~~~~~ & ~~~~~~{\bf 0.1}~~~~~~ & ~~~~~~{\bf 0.2}~~~~~~ & ~~~~~~{\bf 0.3}~~~~~~ & ~~~~~~{\bf 0.4}~~~~~~ & ~~~~~~{\bf 0.5}~~~~~~  \\ \hline
(7,4)-CRGC & 0.02\% & 0.26\% & 3.28\% & 12.53\% & 28.92\% & 50.00\% \\
(7,4)-SRC  & 0.08\% & 0.65\% & 5.24\% & 16.42\% & 33.91\% & 54.70\% \\ \midrule
(7,3)-CRGC & 0.00\% & 0.02\% & 0.45\% & 2.83\% & 9.55\% & 22.53\% \\
(7,3)-SRC  & 0.00\% & 0.06\% & 0.94\% & 4.49\% & 12.87\% & 27.28\% \\ \midrule
(15,5)-CRGC& 0.00\% & 0.00\% & 0.00\% & 0.06\% & 0.89\% & 5.79\% \\
(15,5)-SRC & 0.00\% & 0.00\% & 0.01\% & 0.22\% & 1.92\% & 9.09\% \\ \bottomrule
\end{tabular}
\vspace{2mm}
\caption{Percentage of unrecoverable objects (lost objects) for CRGC and SRC codes in a system with $N=1,000$ nodes. The
values are expressed as a function of the fraction of failed nodes $\Theta$.}
\label{t:lost}
\end{table}

We now evaluate the repair performance when a fraction $\Theta$ of nodes simultaneously fail, where more than one
fragment per object may be lost. We thus study Collaborative Regenerating Codes (CRGC), which have better performance in
terms of both bandwidth utilization and repair latency in comparison to regular RGC. For CRGC we set the repair
parameter $f$ to the number of failed fragments, and to minimize repair traffic we maximize the repair degree, $d=n-f$.
We adapt CRGC's parameters dynamically to the number of failed fragments, so that the evaluation shows the best one
could possibly achieve with CRGC. In Table~\ref{t:lost} we depict for both codes the percentage of lost objects as a
function of the fraction of failed nodes. We can see how the number of lost objects is increased either when we decrease
the redundancy ($n/k$), or when we increase the failure probability $\Theta$. We also note that the values on this
table depend on the total number $N$ of nodes and complete the single-object analysis done in Figure~\ref{f:prob}.

\begin{figure*}
  \centering
  \mbox{\hspace{-.5cm}\includegraphics[scale=0.7]{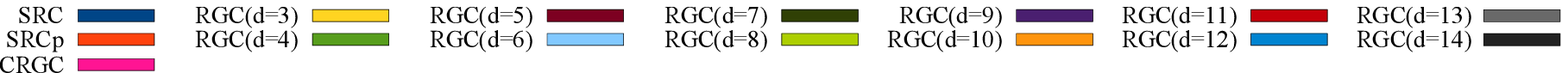}}\vspace{-.4cm}
  \subfloat[Avg. repair time for (7,4).]{\label{fig:corr:t4}\includegraphics[scale=1]{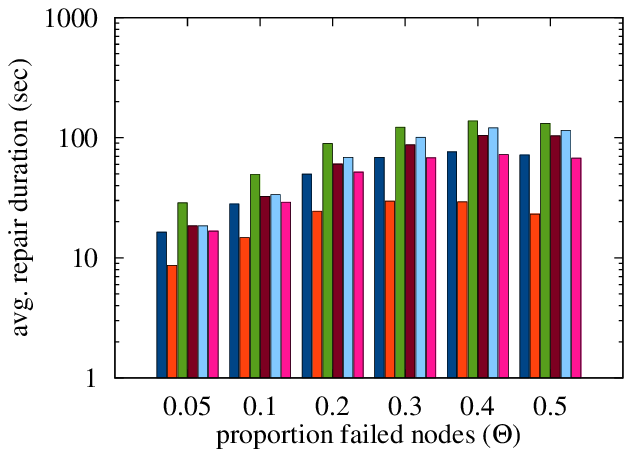}}
  \subfloat[Avg. traffic for (7,4).]{\label{fig:corr:bw4}\includegraphics[scale=1]{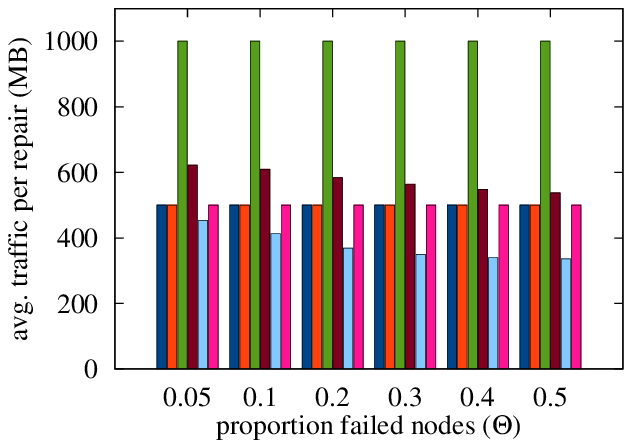}}
  \quad
  \subfloat[Avg. repair time for (7,3).]{\label{fig:corr:t3}\includegraphics[scale=1]{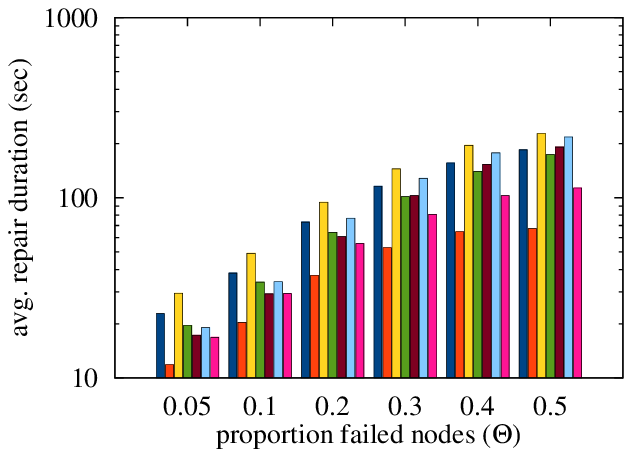}}
  \subfloat[Avg. traffic for (7,3).]{\label{fig:corr:bw3}\includegraphics[scale=1]{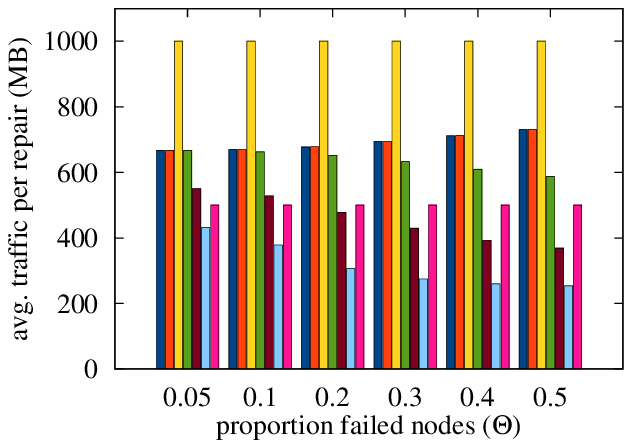}}
  \quad
  \subfloat[Avg. repair time for (15,5).]{\label{fig:corr:t5}\includegraphics[scale=1]{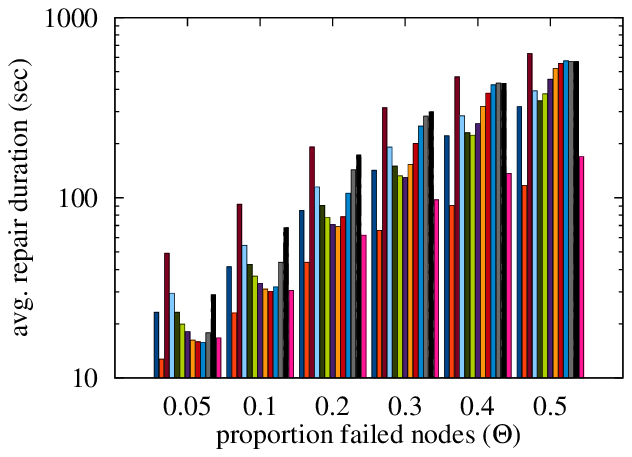}}
  \subfloat[Avg. traffic for (15,5).]{\label{fig:corr:bw5}\includegraphics[scale=1]{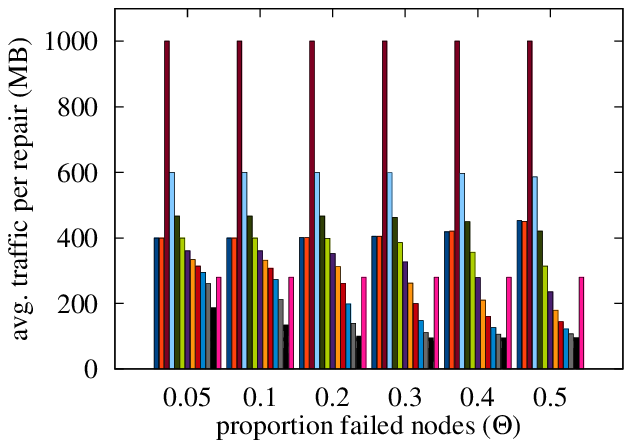}}
  \caption{Analysis of the system performance using different codes when a fraction $\Theta$ of nodes fails
  simultaneously. The size of the objects is $B=1$GB, $L=10,000$ objects are randomly stored in $N=1,000$ nodes.}
  \label{fig:corr}
\end{figure*}

\vspace{.3cm}
Figure~\ref{fig:corr} depicts the results of the multiple failure experiments. Figures~\ref{fig:corr:bw4},
\ref{fig:corr:bw3} and \ref{fig:corr:bw5} display the \emph{average repair traffic} per repaired fragment. We see how in
the (7,3) configuration the SRC/SRCp repair traffic increases with $\Theta$. This happens because of the massive node
failures, which prevent some processes to find any suitable pair of fragments to repair, forcing them to reconstruct the
whole object. The same happens with RGC when less than $d$ fragments survive the failures. The repair process then acts
as a classical EC lazy repair scheme: one of the repairs reconstructs the original object and sends new fragments to the
$f-1$ others, which as depicted in \ref{fig:corr:bw4} and \ref{fig:corr:bw3}, can reduce the traffic for some RGC cases.
Finally, note that the classical EC lazy repair technique, RGC($d=k$), does not achieve further traffic savings.

\begin{figure*}[t]
  \centering
  \subfloat[(7,4) code.]{\hspace{-.4cm}\label{fig:bwcharsingle:74}\includegraphics[scale=1]{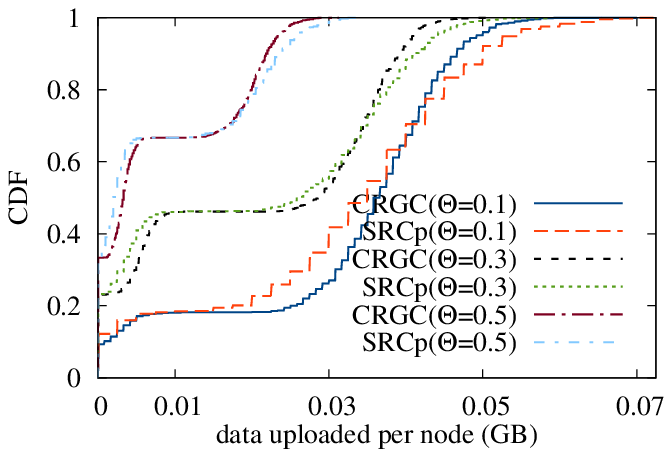}}
  \subfloat[(7,3) code.]{\label{fig:bwcharsingle:73}\hspace{-.8cm}\includegraphics[scale=1]{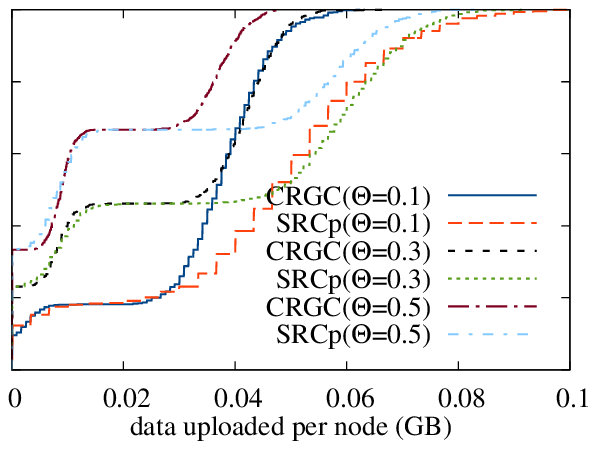}}
  \quad
  \subfloat[(15,5) code.]{\label{fig:bwcahrsingle:155}\hspace{-.8cm}\includegraphics[scale=1]{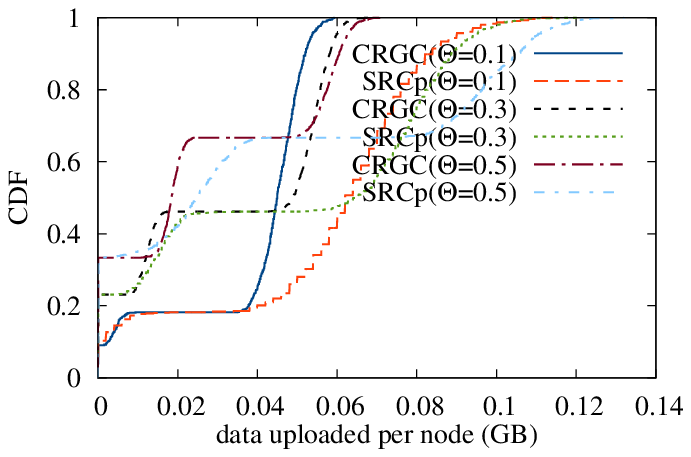}}
  \caption{CDF of the amount of data each node uploads in order to repair a fraction $\Theta$ of failed nodes. Results are obtained
  for a configuration with $L=10,000$, $N=1,000$ and $B=1$GB.}
  \label{fig:bwcharcorr}
\end{figure*}

\vspace{.3cm}
Figures~\ref{fig:corr:t4}, \ref{fig:corr:t3} and \ref{fig:corr:t5} depict the \emph{average fragment repair time} for
different failure probabilities $\Theta$. The first thing that we notice is that in this scenario SRC takes more than
twice the repair time required by SRCp. As noted in Section~\ref{s:single_failure}, this difference is caused by the
random selection of the pair fragments used to repair. We notice how, as the network becomes more saturated --larger
$\Theta$ values--, the gap between SRC and SRCp increases. Furthermore, it may be baffling at first to observe that for
the low redundancy configuration (7,4), repair times decrease when $\Theta=0.5$. This is in fact due to the high number
of lost fragments that cannot be repaired at all for $\Theta=0.5$ (see Table~\ref{t:lost}), which reduces the network
utilization, and hence, reduces the repair times of those repairable objects. Note that for this large $\Theta$ neither
RGC (for large $d$) nor CRGC can repair all objects with the standard repair procedure, and instead have to rely on
reconstructing the entire object and regenerating the missing fragments from it.

\vspace{.3cm}
Finally, similar to Figure~\ref{fig:bwcharsingle}, Figure~\ref{fig:bwcharcorr} illustrates the CDF for the amount of
data each node uploads to repair a fraction $\Theta$ of failed nodes. We only focus on those two configurations that
achieve the best performance in terms of network traffic and upload times: CRGC and SRCp.  As compared to the single
failure case, we can see how the traffic to repair multiple failures is not evenly distributed among all nodes. This is
due to the fact that a fraction $\Theta$ of nodes have no data at all to upload. Besides that, we can also appreciate
how for (7,3) and (15,5) codes the flexibility of CRGC to use any $d$-subset of live nodes in the repair process makes
the overall repair more evenly distributed than for SRCp --steeper CDF curves. Evenly distributed repair traffic avoids
network congestions and explains why CRGC, despite needing to contact more nodes than SRCp, can in some cases --i.e.,
for (15,5) code-- repair faster than SRCp.

\section{Conclusions}
\label{sec:conclude}

In this paper we empirically studied and compared the repair performance of novel codes -- RGC/CRGC and SRC/SRCp --
tailor-made for distributed storage in realistic settings. We found that for single node failures, in most scenarios,
RGC have the largest reduction of repair communication (in fact, the more live nodes contacted $d\geq k$,
the larger the reduction), while SRC/SRCp have slightly less reduction, but significantly better performance than
traditional erasure codes. We also analyzed the effects of data granularity and data placement, showing that although
granularity has no impact on repair performance, highly clustered placements can significantly increase repair times,
compromising data reliability. When multiple nodes fail simultaneously, for certain number of failures, CRGC again has
better traffic reduction than SRC/SRCp, but if the number of failures is very high, then our results confirm that the
traditional erasure code approach of the whole object reconstruction to restore the lost fragments is most efficient in
terms of communication costs, as previously predicted in \cite{OD}.

\vspace{.3cm}
Now in terms of repair time, our new introduced SRCp clearly outperforms all other codes in all studied scenarios. In
fact, in overloaded environments, the performance gain is even more in absence of any well crafted scheduling
mechanisms, i.e., using random suitable nodes. We would also like to note that many of the results for RGC, and all the
results of CRGC assume the existence of codes which can support arbitrary and dynamic choices of $d$ and $f$, and hence
determine what is the best that may be achievable using such codes - if and when they are designed. Thus, all things
considered, we conclude that the pipelined SRC codes have most practical benefit - it is \emph{simple}, has
\emph{similar bandwidth consumption as SRC} (which is typically much lower than traditional erasure codes, but slightly
more than RGC/CRGC) while providing \emph{significant repair speedup} in diverse environments.

\section*{Acknowledgments}
The work of L. Pamies-Juarez and F. Oggier for this paper is supported by the Singapore National Research Foundation
under Research Grant NRF-CRP2-2007-03. A. Datta’s work in this paper has been funded in part by NTU/MoEs AcRF Tier-1
Grant number RG 29/09

\bibliographystyle{plain}
\bibliography{cites}
\end{document}